\documentclass[thmsa,a4paper,11pt]{article}
\usepackage{amssymb}
\usepackage{amsmath}
\usepackage{longtable}
\usepackage{caption}
\usepackage{amsfonts}
\usepackage{graphicx}

\input{tcilatex}

\begin{document}

\author{Dirk Veestraeten\thanks{%
Address correspondence to Dirk Veestraeten, Amsterdam School of Economics,
University of Amsterdam, Roetersstraat 11, 1018 WB Amsterdam, The
Netherlands, e-mail: d.j.m.veestraeten@uva.nl.} \\
University of Amsterdam\\
Amsterdam School of Economics\\
Roetersstraat 11\\
1018 WB Amsterdam\\
The Netherlands\\
\bigskip\bigskip}
\title{On the inverse transform of Laplace transforms that contain (products
of) the parabolic cylinder function\\
\bigskip\bigskip}
\maketitle

\begin{abstract}
\noindent The Laplace transforms of the transition probability density and
distribution functions for the Ornstein-Uhlenbeck process contain the
product of two parabolic cylinder functions, namely $D_{v}\left( x\right)
D_{v}\left( y\right) $ and $D_{v}\left( x\right) D_{v-1}\left( y\right) $,
respectively. The inverse transforms of these products have as yet not been
documented. However, the transition density and distribution functions can
be obtained by alternatively applying Doob's transform to the Kolmogorov
equation and casting the problem in terms of Brownian motion. Linking the
resulting transition density and distribution functions to their Laplace
transforms then specifies the inverse transforms to the aforementioned
products of parabolic cylinder functions. These two results, the recurrence
relation of the parabolic cylinder function and the properties of the
Laplace transform then enable the calculation of inverse transforms also for
countless other combinations in the orders of the parabolic cylinder
functions such as $D_{v}\left( x\right) D_{v-2}\left( y\right) $, $%
D_{v+1}\left( x\right) D_{v-1}\left( y\right) $ and $D_{v}\left( x\right)
D_{v-3}\left( y\right) $.

\vspace{2cm}


\medskip


\newpage
\end{abstract}

\setlength{\baselineskip}{1.25\baselineskip}

\section{Introduction}

The parabolic cylinder function is intensively used in, for instance,
chemical physics (Kalmykov, Coffey and Waldron \cite{kcw96}), lattice field
theory (deLyra, Foong and Gallivant \cite{dfg92}) and astrophysics
(Zaqarashvili and Murawski \cite{zm07}). This paper obtains the inverse
Laplace transforms for products of two parabolic cylinder functions by
exploiting the link between the parabolic cylinder function and the
transition probabilities of the Ornstein-Uhlenbeck process that is used in,
for example, astrophysics (Chandrasekhar \cite{c43}), neurophysiology (Gluss 
\cite{g67}) and finance (Vasicek \cite{v77}).

The Laplace transforms of the transition probability density function -- in
short, the transition density -- and the transition probability distribution
function -- in short, the transition distribution -- of the
Ornstein-Uhlenbeck process contain products of two parabolic cylinder
functions for which the Laplace-transform parameter appears in the order of
both functions. Their inverse transforms have as yet not been documented.
However, the transition density and the transition distribution of the
Ornstein-Uhlenbeck process can be obtained by applying Doob's transform \cite%
{d42} to the Kolmogorov equation (Cox and Miller \cite{cm72}, Risken \cite%
{r89}) in view of casting the latter equation in terms of Brownian motion
for which solutions are well-documented. Combining the resulting transition
density and transition distribution with their Laplace transforms then
yields the desired inverse transforms for Laplace transforms that contain,
in stylised form, $D_{v}\left( x\right) D_{v}\left( y\right) $ (transition
density) and $D_{v}\left( x\right) D_{v-1}\left( y\right) $ (transition
distribution). The latter two results, the recurrence relation for the
parabolic cylinder function and the properties of the Laplace transform then
also enable the calculation of inverse transforms for countless other
combinations in the orders of the parabolic cylinder functions, such as $%
D_{v}\left( x\right) D_{v-2}\left( y\right) $, $D_{v+1}\left( x\right)
D_{v-1}\left( y\right) $ and $D_{v}\left( x\right) D_{v-3}\left( y\right) .$
These results are new to the literature as the inverse Laplace transforms
for products of two parabolic cylinder functions in Erd\'{e}lyi et al. \cite%
{emoti154} and Prudnikov, Brychkov and Marichev \cite{pbm592} do not allow
the Laplace-transform parameter to emerge in the order of the parabolic
cylinder functions.

The inverse Laplace transforms that are derived in this paper can be
specialised in interesting directions. For instance, employing $x=y$ can
yield expressions for the inverse transform of squared parabolic cylinder
functions. Setting either $x=0$ or $y=0$ returns expressions for single
parabolic cylinder functions and the limiting case of $x=y=0$ produces
inverse transforms for ratios of gamma functions.

The remainder of the paper is organised as follows. Section $2$ lists the
properties of the Laplace transform and the parabolic cylinder function that
will be used in the paper. Section $3$ derives the Laplace transforms and
the transition density and transition distribution for the
Ornstein-Uhlenbeck process. Section $4$ illustrates how the resulting two
inverse Laplace transforms can be extended to other orders. Section $5$
presents four specialisations for single parabolic cylinder functions and
ratios of gamma functions and discusses their relation to expressions that
are documented in the literature.

\section{Properties of the Laplace transform and the parabolic cylinder
function}

Extensive detail on the Laplace transform and the parabolic cylinder
function can be found in, for instance, Debnath and Bhatta \cite{db07} and
Erd\'{e}lyi et al. \cite{emoth253}, respectively. This section therefore
only lists the properties that are employed in deriving the below inverse
Laplace transforms.

\subsection{The Laplace transform and its properties}

The Laplace transform of the original function $f\left( t\right) $ is%
\begin{equation*}
L\left\{ f\left( t\right) \right\} =\overline{f}\left( s\right) =\dint
\limits_{0}^{+\infty}\exp\left( -st\right) f\left( t\right) dt, 
\end{equation*}
\noindent where $\func{Re}s>0$. The inverse Laplace transform is denoted by%
\begin{equation*}
L^{-1}\left\{ \overline{f}\left( s\right) \right\} =f\left( t\right) . 
\end{equation*}
\noindent The linearity property of the Laplace transform gives%
\begin{equation}
L\left\{ af\left( t\right) +bg\left( t\right) \right\} =a\overline {f}\left(
s\right) +b\overline{g}\left( s\right) .   \tag{2.1}  \label{ltlin1}
\end{equation}
\noindent Differentiation of the original function $f\left( t\right) $ yields%
\begin{equation}
L\left\{ f^{\prime}\left( t\right) \right\} =s\overline{f}\left( s\right)
-f\left( 0\right)   \tag{2.2}  \label{ltdiff}
\end{equation}
\noindent and linear transformation in the image function gives%
\begin{equation}
L\left\{ \exp\left( at\right) f\left( t\right) \right\} =\overline {f}\left(
s-a\right) .   \tag{2.3}  \label{ltlin2}
\end{equation}

\subsection{The parabolic cylinder function and its properties}

The parabolic cylinder function of order $v$ with argument $z$ is denoted by 
$D_{v}\left( z\right) $. Equation ($8.2.9$) in Erd\'{e}lyi et al. \cite%
{emoth253} specifies the parabolic cylinder function for $v=0$ as%
\begin{equation}
D_{0}\left( z\right) =\exp\left( -\dfrac{z^{2}}{4}\right) .   \tag{2.4}
\label{d0}
\end{equation}
\noindent Equations ($19.3.1$) and ($19.3.5$) in Abramowitz and Stegun \cite%
{as72} give%
\begin{equation}
D_{v}\left( 0\right) =\dfrac{2^{v/2}\sqrt{\pi}}{\Gamma\left( \dfrac{1-v}{2}%
\right) },   \tag{2.5}  \label{pcf0}
\end{equation}
\noindent where $\Gamma\left( v\right) $ denotes the gamma function.
Simplifications in the below results often employ the recurrence and the
duplication formulas for the gamma function%
\begin{align}
& \Gamma\left( v+1\right) =v\Gamma\left( v\right) ,   \tag{2.6a}
\label{gam1} \\
& \Gamma\left( 2v\right) =\dfrac{1}{\sqrt{2\pi}}2^{2v-\frac{1}{2}%
}\Gamma\left( v\right) \Gamma\left( v+\frac{1}{2}\right) ,   \tag{2.6b}
\label{gam2}
\end{align}
\noindent see Equations ($6.1.15$) and ($6.1.18$) in Abramowitz and Stegun 
\cite{as72}.

The derivative of $D_{v}\left( z\right) $ to its argument is given by%
\begin{align}
& D_{v}^{\prime}\left( z\right) =\dfrac{1}{2}zD_{v}\left( z\right)
-D_{v+1}\left( z\right) ,  \tag{2.7a}  \label{deriv1} \\
& D_{v}^{\prime}\left( -z\right) =\dfrac{1}{2}zD_{v}\left( -z\right)
+D_{v+1}\left( -z\right) ,   \tag{2.7b}  \label{deriv2}
\end{align}
\noindent as can be seen from Equation ($8.2.16$) in Erd\'{e}lyi et al. \cite%
{emoth253}. Equation ($8.2.10$) of Erd\'{e}lyi et al. \cite{emoth253}
specifies the Wronskian as%
\begin{equation}
D_{v}\left( z\right) D_{v}^{\prime}\left( -z\right) -D_{v}\left( -z\right)
D_{v}^{\prime}\left( z\right) =\dfrac{\sqrt{2\pi}}{\Gamma\left( -v\right) }. 
\tag{2.8}  \label{wronskian}
\end{equation}
\noindent The recurrence relation for the parabolic cylinder function is
given in Equation ($8.2.14$) of Erd\'{e}lyi et al. \cite{emoth253}%
\begin{equation}
D_{v+1}\left( z\right) -zD_{v}\left( z\right) +vD_{v-1}\left( z\right) =0.  
\tag{2.9}  \label{recurrence}
\end{equation}
\noindent The below identity directly follows from Equations (\ref{deriv1}%
)--(\ref{recurrence})%
\begin{equation}
D_{v}\left( z\right) D_{v-1}\left( -z\right) +D_{v}\left( -z\right)
D_{v-1}\left( z\right) =\dfrac{\sqrt{2\pi}}{-v\Gamma\left( -v\right) }.  
\tag{2.10}  \label{identity}
\end{equation}
\noindent Also the following two integral expressions can be obtained from
the above properties%
\begin{align}
& \dint \exp\left( -\frac{z^{2}}{4}\right) D_{v}\left( z\right)
dz=-\exp\left( -\frac{z^{2}}{4}\right) D_{v-1}\left( z\right) ,  \tag{2.11a}
\label{int1} \\
& \dint \exp\left( -\frac{z^{2}}{4}\right) D_{v}\left( -z\right)
dz=\exp\left( -\frac{z^{2}}{4}\right) D_{v-1}\left( -z\right) .   \tag{2.11b}
\label{int2}
\end{align}

\section{The transition density and transition distribution of the
Ornstein-Uhlenbeck process and their (inverse) Laplace transform}

Let $W=\{W_{t},$ $t\geqslant0\}$ be an Ornstein-Uhlenbeck process with
initial value $W_{0}=w_{0}$. Its dynamics are given by%
\begin{equation}
dW_{t}=\left( \alpha-\beta W_{t}\right) dt+\sigma dZ_{t}\text{,}   \tag{3.1}
\label{ou}
\end{equation}
\noindent with $\beta>0$ and $t>0$. The positive value of $\beta$ implies
that this stochastic process reverts to its long-term mean that is at $%
\tfrac{\alpha}{\beta}$. The instantaneous variance is given by $\sigma^{2} $%
, with $\sigma>0$, and $dZ_{t}$ is the increment of a Wiener process.

The transition distribution is defined as $P\left( w,t|w_{0}\right)
=\Pr\left\{ W\left( t\right) \leqslant w|W\left( 0\right) =w_{0}\right\} $
and the transition density then is $p\left( w,t|w_{0}\right) =\dfrac{\partial%
}{\partial w}P\left( w,t|w_{0}\right) $. The transition density specifies
the probability of attaining $w$ at time $t$ given that the process
initially is at the source point $w_{0}$. The transition density satisfies
the Kolmogorov forward and backward equations as discussed in, for instance,
Cox and Miller \cite{cm72} and Risken \cite{r89}. The Kolmogorov backward
equation for the process in Equation (\ref{ou}) is%
\begin{equation*}
\frac{1}{2}\sigma^{2}\frac{\partial^{2}p\left( w,t|w_{0}\right) }{\partial
w_{0}^{2}}+\left( \alpha-\beta w_{0}\right) \frac{\partial p\left(
w,t|w_{0}\right) }{\partial w_{0}}=\frac{\partial p\left( w,t|w_{0}\right) }{%
\partial t}. 
\end{equation*}
\noindent The initial condition is given by%
\begin{equation*}
p\left( w,t|w_{0}\right) =\delta\left( w-w_{0}\right) \qquad\text{for }t=0, 
\end{equation*}
\noindent where $\delta\left( \cdot\right) $ is the Dirac delta function
that guarantees that all initial probability mass is located on the initial
state of the process.

The Laplace transform of the transition density, $\overline{p}\left(
w,s|w_{0}\right) $, is%
\begin{equation*}
\overline{p}\left( w,s|w_{0}\right) =\dint \limits_{0}^{+\infty}\exp\left(
-st\right) p\left( w,t|w_{0}\right) dt, 
\end{equation*}
\noindent with $\func{Re}s>0$. The Kolmogorov backward equation then can be
rewritten as%
\begin{equation}
\frac{1}{2}\sigma^{2}\dfrac{d^{2}\overline{p}\left( w,s|w_{0}\right) }{%
dw_{0}^{2}}+\left( \alpha-\beta w_{0}\right) \frac{d\overline{p}\left(
w,s|w_{0}\right) }{dw_{0}}-s\hspace{0.05cm}\overline{p}\left(
w,s|w_{0}\right) =-\delta\left( w-w_{0}\right) \text{.}   \tag{3.2}
\label{ltkbe}
\end{equation}
\noindent The substitutions $z_{0}=w_{0}-\tfrac{\alpha}{\beta}$ and $q_{0}=%
\tfrac{\sqrt{2\beta}}{\sigma}z_{0}$, and the transformation $\overline{v}%
\left( q,s|q_{0}\right) =\exp\left( -\tfrac{q_{0}^{2}}{4}\right) \overline{p}%
\left( q,s|q_{0}\right) $ simplify the homogenous part of Equation (\ref%
{ltkbe}) into%
\begin{equation*}
\dfrac{d^{2}\overline{v}\left( q,s|q_{0}\right) }{dq_{0}^{2}}+\left\{ \dfrac{%
1}{2}-\dfrac{s}{\beta}-\dfrac{q_{0}^{2}}{4}\right\} \overline {v}\left(
q,s|q_{0}\right) =0. 
\end{equation*}
\noindent The parabolic cylinder functions $D_{-s/\beta}\left( q_{0}\right) $
and $D_{-s/\beta}\left( -q_{0}\right) $ are two linearly independent
solutions to the latter equation (see Equation ($8.2.5$) in Erd\'{e}lyi et
al. \cite{emoth253}). The resulting Green's function, $G\left(
w,w_{0}\right) $, is%
\begin{align*}
& G\left( w,w_{0}\right) =\left. G\left( w,w_{0}\right) \right\vert
_{-\infty\leqslant w_{0}\leqslant w}+\left. G\left( w,w_{0}\right)
\right\vert _{w\leqslant w_{0}\leqslant+\infty}\text{,} \\
& \text{with }\left. G\left( w,w_{0}\right) \right\vert _{-\infty\leqslant
w_{0}\leqslant w}=A\exp\left( \dfrac{\left( \beta w_{0}-\alpha\right) ^{2}}{%
2\sigma^{2}\beta}\right) D_{-s/\beta}\left( -\dfrac{\sqrt{2}\left( \beta
w_{0}-\alpha\right) }{\sigma\sqrt{\beta}}\right) , \\
& \hspace{0.85cm}\left. G\left( w,w_{0}\right) \right\vert _{w\leqslant
w_{0}\leqslant+\infty}=B\exp\left( \dfrac{\left( \beta w_{0}-\alpha\right)
^{2}}{2\sigma^{2}\beta}\right) D_{-s/\beta}\left( \dfrac{\sqrt{2}\left(
\beta w_{0}-\alpha\right) }{\sigma\sqrt{\beta}}\right) ,
\end{align*}
\noindent for which the constants $A$ and $B$ are to be obtained from the
conditions%
\begin{align*}
& \left. G\left( w,w_{0}\right) \right\vert _{w\leqslant
w_{0}\leqslant+\infty}-\left. G\left( w,w_{0}\right) \right\vert _{-\infty
\leqslant w_{0}\leqslant w}=0\text{ \hspace{1.5cm}for }w_{0}=w\text{,} \\
& \dfrac{\left. dG\left( w,w_{0}\right) \right\vert _{w\leqslant
w_{0}\leqslant+\infty}}{dw_{0}}-\dfrac{\left. dG\left( w,w_{0}\right)
\right\vert _{-\infty\leqslant w_{0}\leqslant w}}{dw_{0}}=-\dfrac{2}{%
\sigma^{2}}\text{ \hspace{0.4cm}for }w_{0}=w.
\end{align*}
\noindent The derivatives and the Wronskian in Equations (\ref{deriv1})--(%
\ref{wronskian}) allow to express the Laplace transform of the transition
density as%
\begin{equation}
\overline{p}\left( w,s|w_{0}\right) =\left\{ 
\begin{array}{l}
\dfrac{\Gamma\left( s/\beta\right) }{\sigma\sqrt{\pi\beta}}\exp\left( \dfrac{%
\left( \beta w_{0}-\alpha\right) ^{2}}{2\sigma^{2}\beta}-\dfrac{\left( \beta
w-\alpha\right) ^{2}}{2\sigma^{2}\beta}\right) \\ 
\ \hspace{0.5cm}\times D_{-s/\beta}\left( \dfrac{\sqrt{2}\left( \beta
w-\alpha\right) }{\sigma\sqrt{\beta}}\right) D_{-s/\beta}\left( -\dfrac{%
\sqrt{2}\left( \beta w_{0}-\alpha\right) }{\sigma\sqrt{\beta}}\right) \text{
for}-\infty\leqslant w_{0}\leqslant w, \\ 
\dfrac{\Gamma\left( s/\beta\right) }{\sigma\sqrt{\pi\beta}}\exp\left( \dfrac{%
\left( \beta w_{0}-\alpha\right) ^{2}}{2\sigma^{2}\beta}-\dfrac{\left( \beta
w-\alpha\right) ^{2}}{2\sigma^{2}\beta}\right) \\ 
\ \hspace{0.5cm}\times D_{-s/\beta}\left( -\dfrac{\sqrt{2}\left( \beta
w-\alpha\right) }{\sigma\sqrt{\beta}}\right) D_{-s/\beta}\left( \dfrac{\sqrt{%
2}\left( \beta w_{0}-\alpha\right) }{\sigma\sqrt{\beta}}\right) \text{ for }%
w\leqslant w_{0}\leqslant+\infty.%
\end{array}
\right.   \tag{3.3}  \label{lttransdens}
\end{equation}
\noindent However, the literature did not obtain the original function $%
p\left( w,t|w_{0}\right) $ by inverting the Laplace transform (\ref%
{lttransdens}). Instead, Doob's transform, see Doob \cite{d42}, offers a
simpler route by simplifying the Kolmogorov forward equation into the
diffusion equation for which solutions are widely available. The resulting
transition density $p\left( w,t|w_{0}\right) $ is documented in, amongst
others, Gluss \cite{g67} and Cox and Miller \cite{cm72}) as%
\begin{equation}
p\left( w,t|w_{0}\right) =\dfrac{\sqrt{\beta}}{\sqrt{\pi\sigma^{2}\left(
1-\exp\left( -2\beta t\right) \right) }}\exp\left( -\dfrac{\left( \left(
\beta w-\alpha\right) -\left( \beta w_{0}-\alpha\right) \exp\left( -\beta
t\right) \right) ^{2}}{\beta\sigma^{2}\left( 1-\exp\left( -2\beta t\right)
\right) }\right) .   \tag{3.4}  \label{transdens}
\end{equation}
\noindent The transition density (\ref{transdens}) thus specifies the
inverse Laplace transform of Equation (\ref{lttransdens}).

The transition distribution, $P\left( w_{1},t|w_{0}\right) $, specifies the
cumulative probability mass that is situated below $w_{1}$. Its Laplace
transform is%
\begin{equation*}
\overline{P}\left( w_{1},s|w_{0}\right) =\dint
\limits_{0}^{+\infty}\exp\left( -st\right) P\left( w_{1},t|w_{0}\right) dt, 
\end{equation*}
\noindent where $\func{Re}s>0$. The Laplace transform $\overline {P}\left(
w_{1},s|w_{0}\right) $ can be obtained from the Laplace transform of the
transition density by employing the definition of the transition
distribution as follows%
\begin{align*}
\overline{P}\left( w_{1},s|w_{0}\right) & =\dint
\limits_{0}^{+\infty}\exp\left( -st\right) \left\{ \dint
\limits_{-\infty}^{w_{1}}p\left( w,t|w_{0}\right) dw\right\} dt \\
\overline{P}\left( w_{1},s|w_{0}\right) & =\dint
\limits_{-\infty}^{w_{1}}\left\{ \dint \limits_{0}^{+\infty}\exp\left(
-st\right) p\left( w,t|w_{0}\right) dt\right\} dw \\
\overline{P}\left( w_{1},s|w_{0}\right) & =\dint \limits_{-\infty}^{w_{1}}%
\overline{p}\left( w,s|w_{0}\right) dw.
\end{align*}
\noindent The expression for the Laplace transform $\overline{P}\left(
w_{1},s|w_{0}\right) $ will differ in function of the location of $w_{1}$
with respect to $w_{0}$. However, the precise position of $w_{1}$ does not
affect the below results such that we continue here with the restriction $%
w_{1}\geqslant w_{0}$. The Laplace transform (\ref{lttransdens}) gives%
\begin{align*}
\overline{P}\left( w_{1},s|w_{0}\right) & =\dfrac{\Gamma\left(
s/\beta\right) }{\sigma\sqrt{\pi\beta}}\exp\left( \dfrac{\left( \beta
w_{0}-\alpha\right) ^{2}}{2\sigma^{2}\beta}\right) D_{-s/\beta}\left( -%
\dfrac{\sqrt{2}\left( \beta w_{0}-\alpha\right) }{\sigma\sqrt{\beta}}\right)
\\
& \hspace{0.5cm}\times\dint \limits_{w_{0}}^{w_{1}}\exp\left( -\dfrac{\left(
\beta w-\alpha\right) ^{2}}{2\sigma^{2}\beta }\right) D_{-s/\beta}\left( 
\dfrac{\sqrt{2}\left( \beta w-\alpha\right) }{\sigma\sqrt{\beta}}\right) dw
\\
& +\dfrac{\Gamma\left( s/\beta\right) }{\sigma\sqrt{\pi\beta}}\exp\left( 
\dfrac{\left( \beta w_{0}-\alpha\right) ^{2}}{2\sigma^{2}\beta}\right)
D_{-s/\beta}\left( \dfrac{\sqrt{2}\left( \beta w_{0}-\alpha\right) }{\sigma%
\sqrt{\beta}}\right) \\
& \hspace{0.5cm}\times\dint \limits_{-\infty}^{w_{0}}\exp\left( -\dfrac{%
\left( \beta w-\alpha\right) ^{2}}{2\sigma^{2}\beta }\right)
D_{-s/\beta}\left( -\dfrac{\sqrt{2}\left( \beta w-\alpha\right) }{\sigma%
\sqrt{\beta}}\right) dw.
\end{align*}
\noindent The integrals (\ref{int1}) and (\ref{int2}) and the identity (\ref%
{identity}) then allow to specify $\overline{P}\left( w_{1},s|w_{0}\right) $%
\ as%
\begin{align}
& \overline{P}\left( w_{1},s|w_{0}\right) =\dfrac{1}{s}-\dfrac{\Gamma\left(
s/\beta\right) }{\beta\sqrt{2\pi}}\exp\left( \dfrac{\left( \beta
w_{0}-\alpha\right) ^{2}}{2\sigma^{2}\beta}-\dfrac{\left( \beta
w_{1}-\alpha\right) ^{2}}{2\sigma^{2}\beta}\right)  \tag{3.5}
\label{lttransdist} \\
& \hspace{2cm}\times D_{-s/\beta}\left( -\dfrac{\sqrt{2}\left( \beta
w_{0}-\alpha\right) }{\sigma\sqrt{\beta}}\right) D_{-1-s/\beta}\left( \dfrac{%
\sqrt{2}\left( \beta w_{1}-\alpha\right) }{\sigma\sqrt{\beta}}\right) \text{
for }w_{1}\geqslant w_{0}.  \notag
\end{align}
\noindent Equation (\ref{lttransdist}) has a clear probabilistic
interpretation as the inverse transform of the term $\dfrac{1}{s}$ is $1$.
The latter term thus corresponds with the cumulative probability of the
Ornstein-Uhlenbeck process over its entire domain, i.e. $\left(
-\infty,+\infty\right) $. The second term then must be seen as representing
the Laplace transform of the loss\ of probability that arises when $w_{1}$
takes on a value below $+\infty$. As required, lifting $w_{1}$ to $+\infty$
reduces the second term to zero as can be seen from Equations ($19.3.1$) and
($19.8.1$) in Abramowitz and Stegun \cite{as72}.

The original function $P\left( w_{1},s|w_{0}\right) $ can directly be
obtained by integrating the transition density (\ref{transdens})%
\begin{equation}
P\left( w_{1},t|w_{0}\right) =\dfrac{1}{2}\text{erfc}\left( -\dfrac{\left(
\left( \beta w_{1}-\alpha\right) -\left( \beta w_{0}-\alpha\right)
\exp\left( -\beta t\right) \right) }{\sqrt{\beta\sigma^{2}\left(
1-\exp\left( -2\beta t\right) \right) }}\right) ,   \tag{3.6}
\label{transdist}
\end{equation}
\noindent where erfc$\left( z\right) $ denotes the complementary error
function, see Abramowitz and Stegun \cite{as72}.

The inverse transform of the Laplace transform (\ref{lttransdist}) thus is
given by the transition distribution (\ref{transdist}).

\section{Inverse transforms of Laplace transforms that contain products of
two parabolic cylinder functions}

This section illustrates how the recurrence relation of the parabolic
cylinder function and the above properties of the Laplace transform can be
used to obtain further results out of the two central inverse Laplace
transforms. However, the below $10$ inverse transforms, that are also
collected in Table $1$, thus must be seen as a subset of the infinitely many
inverse transforms that can be obtained.%
\begin{equation*}
\fbox{Table 1: around here.}
\end{equation*}
\noindent We start with the (inverse) Laplace transform of the transition
density in Equations (\ref{lttransdens}) and (\ref{transdens}) for $%
-\infty\leqslant w_{0}\leqslant w$. Notation is simplified by using $x=-%
\tfrac{\sqrt{2}\left( \beta w_{0}-\alpha\right) }{\sigma\sqrt{\beta}}$ and $%
y=\tfrac{\sqrt{2}\left( \beta w-\alpha\right) }{\sigma\sqrt{\beta}}$. The
linearity property (\ref{ltlin1}) allows to re-express the Laplace transform
in Equation (\ref{lttransdens}) as $\overline{f}\left( s\right)
=\Gamma\left( s/\beta\right) D_{-s/\beta}\left( x\right) D_{-s/\beta }\left(
y\right) $ for which the previous condition $-\infty\leqslant w_{0}\leqslant
w$ translates into $x+y\geqslant0$. Note that the latter condition does not
impose any additional (linear) relation between the two arguments. Linearly
transforming the image function via Equation (\ref{ltlin2}) for $s+c$ with $%
c\geqslant0$ creates a more versatile specification for the order of the
parabolic cylinder functions%
\begin{align}
& L^{-1}\left\{ \Gamma\left( \tfrac{s+c}{\beta}\right) D_{-\left( s+c\right)
/\beta}\left( x\right) D_{-\left( s+c\right) /\beta}\left( y\right) \right\}
=  \notag \\
& \hspace{0.5cm}=\dfrac{\beta\exp\left( -ct\right) }{\sqrt{1-\exp\left(
-2\beta t\right) }}\exp\left( \dfrac{y^{2}-x^{2}}{4}\right) \exp\left( -%
\dfrac{\left( y+x\exp\left( -\beta t\right) \right) ^{2}}{2\left(
1-\exp\left( -2\beta t\right) \right) }\right)  \notag \\
& \hspace{0.5cm}\left[ \func{Re}s>0,\beta>0,c\geqslant0,x+y\geqslant 0\right]
.   \tag{4.1}  \label{4.1}
\end{align}
\noindent It is to be noted that Equation ($4.1$), with the same condition,
also arises when evaluating the Laplace transform in Equation (\ref%
{lttransdens}) for the domain $w\leqslant w_{0}\leqslant+\infty$.

Applying the same steps to the (inverse) Laplace transform for the
transition distribution in Equations (\ref{lttransdist}) and (\ref{transdist}%
) gives 
\begin{align}
& L^{-1}\left\{ \Gamma\left( \tfrac{s+c}{\beta}\right) D_{-\left( s+c\right)
/\beta}\left( x\right) D_{-1-\left( s+c\right) /\beta}\left( y\right)
\right\}  \notag \\
& \hspace{0.5cm}=\dfrac{\beta\sqrt{\pi}\exp\left( -ct\right) }{\sqrt{2}}%
\exp\left( \dfrac{y^{2}-x^{2}}{4}\right) \text{erfc}\left( \dfrac {%
y+x\exp\left( -\beta t\right) }{\sqrt{2\left( 1-\exp\left( -2\beta t\right)
\right) }}\right)  \notag \\
& \hspace{0.5cm}\left[ \func{Re}s>0,\beta>0,c\geqslant0,x+y\geqslant 0\right]
.   \tag{4.2}  \label{4.2}
\end{align}
\noindent Fixing the order of the first parabolic cylinder function\ in the
product at $-\left( s+c\right) /\beta$ and using the recurrence relation (%
\ref{recurrence}) on the second parabolic cylinder function with $%
v=-1-\left( s+c\right) /\beta$ gives%
\begin{align}
& \tfrac{s+\beta+c}{\beta}\Gamma\left( \tfrac{s+c}{\beta}\right) D_{-\left(
s+c\right) /\beta}\left( x\right) D_{-2-\left( s+c\right) /\beta}\left(
y\right) =  \tag{4.3}  \label{4.3} \\
& \hspace{0.5cm}-y\Gamma\left( \tfrac{s+c}{\beta}\right) D_{-\left(
s+c\right) /\beta}\left( x\right) D_{-1-\left( s+c\right) /\beta}\left(
y\right) +\Gamma\left( \tfrac{s+c}{\beta}\right) D_{-\left( s+c\right)
/\beta}\left( x\right) D_{-\left( s+c\right) /\beta}\left( y\right) .  \notag
\end{align}
\noindent Plugging the inverse Laplace transforms (\ref{4.1}) and (\ref{4.2}%
) into relation (\ref{4.3}) gives%
\begin{align}
& L^{-1}\left\{ \tfrac{s+\beta+c}{\beta}\Gamma\left( \tfrac{s+c}{\beta }%
\right) D_{-\left( s+c\right) /\beta}\left( x\right) D_{-2-\left( s+c\right)
/\beta}\left( y\right) \right\}  \notag \\
& \hspace{0.5cm}=\exp\left( \dfrac{y^{2}-x^{2}}{4}\right) \left\{ \dfrac{%
\beta\exp\left( -ct\right) }{\sqrt{1-\exp\left( -2\beta t\right) }}%
\exp\left( -\dfrac{\left( y+x\exp\left( -\beta t\right) \right) ^{2}}{%
2\left( 1-\exp\left( -2\beta t\right) \right) }\right) \right.  \notag \\
& \hspace{0.5cm}\left. -\dfrac{y\beta\sqrt{\pi}\exp\left( -ct\right) }{\sqrt{%
2}}\text{erfc}\left( \dfrac{y+x\exp\left( -\beta t\right) }{\sqrt{2\left(
1-\exp\left( -2\beta t\right) \right) }}\right) \right\}  \notag \\
& \hspace{0.5cm}\left[ \func{Re}s>0,\beta>0,c\geqslant0,x+y\geqslant 0\right]
.   \tag{4.4}  \label{4.4}
\end{align}
\noindent Increasing $v$ to $-\left( s+c\right) /\beta$ in the recurrence
relation (\ref{recurrence}) yields%
\begin{align}
& \Gamma\left( \tfrac{s+c}{\beta}\right) D_{-\left( s+c\right) /\beta
}\left( x\right) D_{1-\left( s+c\right) /\beta}\left( y\right) =  \tag{4.5}
\label{4.5} \\
& \hspace{0.5cm}y\Gamma\left( \tfrac{s+c}{\beta}\right) D_{-\left(
s+c\right) /\beta}\left( x\right) D_{-\left( s+c\right) /\beta}\left(
y\right) +\tfrac{s+c}{\beta}\Gamma\left( \tfrac{s+c}{\beta}\right)
D_{-\left( s+c\right) /\beta}\left( x\right) D_{-1-\left( s+c\right)
/\beta}\left( y\right) .  \notag
\end{align}
\noindent The inverse transform of $s\Gamma\left( \tfrac{s+c}{\beta}\right)
D_{-\left( s+c\right) /\beta}\left( x\right) D_{-1-\left( s+c\right)
/\beta}\left( y\right) $ is obtained by applying the differentiation
property (\ref{ltdiff}) to the inverse transform (\ref{4.2}). The term $%
f\left( 0\right) $ is $\tfrac{\beta\sqrt{\pi}}{\sqrt{2}}$ for $x+y=0$ and $0$
otherwise, for which we employ the notation $\left. \tfrac{\beta\sqrt {\pi}}{%
\sqrt{2}}\right\vert _{x+y=0}$. Indeed, the complementary error function for 
$x+y\geqslant0$ and $t=0$ gives erfc$\left( +\infty\right) =0$, whereas it
yields erfc$\left( 0\right) =1$ for $x+y=0$. The inverse transform then can
be written as%
\begin{align}
& L^{-1}\left\{ s\Gamma\left( \tfrac{s+c}{\beta}\right) D_{-\left(
s+c\right) /\beta}\left( x\right) D_{-1-\left( s+c\right) /\beta}\left(
y\right) -\left. \tfrac{\beta\sqrt{\pi}}{\sqrt{2}}\right\vert
_{x+y=0}\right\}  \notag \\
& \hspace{0.5cm}=\exp\left( \dfrac{y^{2}-x^{2}}{4}\right) \left\{ \dfrac{%
\beta^{2}\exp\left( -\left( \beta+c\right) t\right) }{\left( 1-\exp\left(
-2\beta t\right) \right) ^{\frac{3}{2}}}\left( x+y\exp\left( -\beta t\right)
\right) \exp\left( -\dfrac{\left( y+x\exp\left( -\beta t\right) \right) ^{2}%
}{2\left( 1-\exp\left( -2\beta t\right) \right) }\right) \right.  \notag \\
& \hspace{0.5cm}\left. -\dfrac{c\beta\sqrt{\pi}\exp\left( -ct\right) }{\sqrt{%
2}}\text{erfc}\left( \dfrac{y+x\exp\left( -\beta t\right) }{\sqrt{2\left(
1-\exp\left( -2\beta t\right) \right) }}\right) \right\}  \notag \\
& \hspace{0.5cm}\left[ \func{Re}s>0,\beta>0,c\geqslant0,x+y\geqslant 0\right]
.   \tag{4.6}  \label{4.6}
\end{align}
\noindent Plugging the inverse transforms (\ref{4.1}), (\ref{4.2}) and (\ref%
{4.6}) into relation (\ref{4.5}) yields%
\begin{align}
& L^{-1}\left\{ \Gamma\left( \tfrac{s+c}{\beta}\right) D_{-\left( s+c\right)
/\beta}\left( x\right) D_{1-\left( s+c\right) /\beta}\left( y\right) -\left. 
\tfrac{\sqrt{\pi}}{\sqrt{2}}\right\vert _{x+y=0}\right\}  \notag \\
& \hspace{0.5cm}=\dfrac{\beta\exp\left( -ct\right) }{\left( 1-\exp\left(
-2\beta t\right) \right) ^{\frac{3}{2}}}\left( y+x\exp\left( -\beta t\right)
\right) \exp\left( \dfrac{y^{2}-x^{2}}{4}\right) \exp\left( -\dfrac{\left(
y+x\exp\left( -\beta t\right) \right) ^{2}}{2\left( 1-\exp\left( -2\beta
t\right) \right) }\right)  \notag \\
& \hspace{0.5cm}\left[ \func{Re}s>0,\beta>0,c\geqslant0,x+y\geqslant 0\right]
.   \tag{4.7}  \label{4.7}
\end{align}
\noindent Fixing the order in the second parabolic cylinder function at $%
-1-\left( s+c\right) /\beta$ and using $v=-\left( s+c\right) /\beta$ within
the recurrence relation (\ref{recurrence}) for the first parabolic cylinder
function gives%
\begin{align}
& \Gamma\left( \tfrac{s+c}{\beta}\right) D_{1-\left( s+c\right) /\beta
}\left( x\right) D_{-1-\left( s+c\right) /\beta}\left( y\right) =  \tag{4.8}
\label{4.8} \\
& \hspace{0.5cm}x\Gamma\left( \tfrac{s+c}{\beta}\right) D_{-\left(
s+c\right) /\beta}\left( x\right) D_{-1-\left( s+c\right) /\beta}\left(
y\right) +\tfrac{s+c}{\beta}\Gamma\left( \tfrac{s+c}{\beta}\right)
D_{-1-\left( s+c\right) /\beta}\left( x\right) D_{-1-\left( s+c\right)
/\beta}\left( y\right) .  \notag
\end{align}
\noindent The inverse transform of $\tfrac{s+c}{\beta}\Gamma\left( \tfrac{s+c%
}{\beta}\right) D_{-1-\left( s+c\right) /\beta}\left( x\right) D_{-1-\left(
s+c\right) /\beta}\left( y\right) $ is obtained via applying the linear
transformation (\ref{ltlin2}) for $\overline{f}\left( s+\beta\right) $ to
Equation (\ref{4.1}) and using the recurrence formula for the gamma function
in Equation (\ref{gam2})%
\begin{align}
& L^{-1}\left\{ \tfrac{s+c}{\beta}\Gamma\left( \tfrac{s+c}{\beta}\right)
D_{-1-\left( s+c\right) /\beta}\left( x\right) D_{-1-\left( s+c\right)
/\beta}\left( y\right) \right\}  \notag \\
& \hspace{0.5cm}=\dfrac{\beta\exp\left( -\left( \beta+c\right) t\right) }{%
\sqrt{1-\exp\left( -2\beta t\right) }}\exp\left( \dfrac{y^{2}-x^{2}}{4}%
\right) \exp\left( -\dfrac{\left( y+x\exp\left( -\beta t\right) \right) ^{2}%
}{2\left( 1-\exp\left( -2\beta t\right) \right) }\right)  \notag \\
& \hspace{0.5cm}\left[ \func{Re}s>0,\beta>0,c\geqslant0,x+y\geqslant 0\right]
.   \tag{4.9}  \label{4.9}
\end{align}
\noindent Connecting the inverse transforms (\ref{4.2}) and (\ref{4.9}) to
the recurrence relation (\ref{4.8}) yields%
\begin{align}
& L^{-1}\left\{ \Gamma\left( \tfrac{s+c}{\beta}\right) D_{1-\left(
s+c\right) /\beta}\left( x\right) D_{-1-\left( s+c\right) /\beta}\left(
y\right) \right\}  \notag \\
& \hspace{0.5cm}=\exp\left( \dfrac{y^{2}-x^{2}}{4}\right) \left\{ \dfrac{%
\beta\exp\left( -\left( \beta+c\right) t\right) }{\sqrt {1-\exp\left(
-2\beta t\right) }}\exp\left( -\dfrac{\left( y+x\exp\left( -\beta t\right)
\right) ^{2}}{2\left( 1-\exp\left( -2\beta t\right) \right) }\right) \right.
\notag \\
& \hspace{0.5cm}\left. +\dfrac{x\beta\sqrt{\pi}\exp\left( -ct\right) }{\sqrt{%
2}}\text{erfc}\left( \dfrac{y+x\exp\left( -\beta t\right) }{\sqrt{2\left(
1-\exp\left( -2\beta t\right) \right) }}\right) \right\}  \notag \\
& \hspace{0.5cm}\left[ \func{Re}s>0,\beta>0,c\geqslant0,x+y\geqslant 0\right]
.   \tag{4.10}  \label{4.10}
\end{align}
\noindent Setting the order in the second parabolic cylinder function at $%
v=-2-\left( s+c\right) /\beta$ and keeping the order of the first one at $%
-\left( s+c\right) /\beta$ gives%
\begin{align}
& \tfrac{s+2\beta+c}{\beta}\Gamma\left( \tfrac{s+c}{\beta}\right) D_{-\left(
s+c\right) /\beta}\left( x\right) D_{-3-\left( s+c\right) /\beta}\left(
y\right) =  \tag{4.11}  \label{4.11} \\
& \hspace{0.5cm}-y\Gamma\left( \tfrac{s+c}{\beta}\right) D_{-\left(
s+c\right) /\beta}\left( x\right) D_{-2-\left( s+c\right) /\beta}\left(
y\right) +\Gamma\left( \tfrac{s+c}{\beta}\right) D_{-\left( s+c\right)
/\beta}\left( x\right) D_{-1-\left( s+c\right) /\beta}\left( y\right) . 
\notag
\end{align}
\noindent The inverse transform for the first term on the right-hand side of
relation (\ref{4.11}) can be obtained from earlier results in two steps.
First, applying the linear transformation in Equation (\ref{ltlin2}) for $%
\overline{f}\left( s+\beta\right) $ and the recurrence formula for the gamma
function (\ref{gam1}) to the inverse transform (\ref{4.10}) gives%
\begin{align}
& L^{-1}\left\{ \tfrac{s+c}{\beta}\Gamma\left( \tfrac{s+c}{\beta}\right)
D_{-\left( s+c\right) /\beta}\left( x\right) D_{-2-\left( s+c\right)
/\beta}\left( y\right) \right\}  \notag \\
& \hspace{0.5cm}=\exp\left( \frac{y^{2}-x^{2}}{4}\right) \left\{ \dfrac{%
\beta\exp\left( -\left( 2\beta+c\right) t\right) }{\sqrt {1-\exp\left(
-2\beta t\right) }}\exp\left( -\dfrac{\left( y+x\exp\left( -\beta t\right)
\right) ^{2}}{2\left( 1-\exp\left( -2\beta t\right) \right) }\right) \right.
\notag \\
& \hspace{0.5cm}\left. +\dfrac{x\beta\sqrt{\pi}\exp\left( -\left(
\beta+c\right) t\right) }{\sqrt{2}}\text{erfc}\left( \dfrac{y+x\exp\left(
-\beta t\right) }{\sqrt{2\left( 1-\exp\left( -2\beta t\right) \right) }}%
\right) \right\}  \notag \\
& \hspace{0.5cm}\left[ \func{Re}s>0,\beta>0,c\geqslant0,x+y\geqslant 0\right]
.   \tag{4.12}  \label{4.12}
\end{align}
\noindent Second, subtracting the latter result from the inverse transform (%
\ref{4.4}) gives%
\begin{align}
& L^{-1}\left\{ \Gamma\left( \tfrac{s+c}{\beta}\right) D_{-\left( s+c\right)
/\beta}\left( x\right) D_{-2-\left( s+c\right) /\beta}\left( y\right)
\right\}  \notag \\
& \hspace{0.5cm}=\exp\left( \frac{y^{2}-x^{2}}{4}\right) \left\{ \beta \sqrt{%
1-\exp\left( -2\beta t\right) }\exp\left( -ct\right) \exp\left( -\dfrac{%
\left( y+x\exp\left( -\beta t\right) \right) ^{2}}{2\left( 1-\exp\left(
-2\beta t\right) \right) }\right) \right.  \notag \\
& \hspace{0.5cm}\left. -\dfrac{\left( x\exp\left( -\beta t\right) +y\right)
\beta\sqrt{\pi}\exp\left( -ct\right) }{\sqrt{2}}\text{erfc}\left( \dfrac{%
y+x\exp\left( -\beta t\right) }{\sqrt{2\left( 1-\exp\left( -2\beta t\right)
\right) }}\right) \right\}  \notag \\
& \hspace{0.5cm}\left[ \func{Re}s>0,\beta>0,c\geqslant0,x+y\geqslant 0\right]
.   \tag{4.13}  \label{4.13}
\end{align}
\noindent Finally, combining the inverse transforms (\ref{4.2}) and (\ref%
{4.13}) with the recurrence relation (\ref{4.11}) yields the desired result%
\begin{align}
& L^{-1}\left\{ \tfrac{s+2\beta+c}{\beta}\Gamma\left( \tfrac{s+c}{\beta }%
\right) D_{-\left( s+c\right) /\beta}\left( x\right) D_{-3-\left( s+c\right)
/\beta}\left( y\right) \right\}  \notag \\
& \hspace{0.5cm}=\exp\left( \frac{y^{2}-x^{2}}{4}\right) \left\{ -y\beta%
\sqrt{1-\exp\left( -2\beta t\right) }\exp\left( -ct\right) \exp\left( -%
\dfrac{\left( y+x\exp\left( -\beta t\right) \right) ^{2}}{2\left(
1-\exp\left( -2\beta t\right) \right) }\right) \right.  \notag \\
& \hspace{0.5cm}\left. +\dfrac{\left( y\left( x\exp\left( -\beta t\right)
+y\right) +1\right) \beta\sqrt{\pi}\exp\left( -ct\right) }{\sqrt{2}}\text{%
erfc}\left( \dfrac{y+x\exp\left( -\beta t\right) }{\sqrt{2\left(
1-\exp\left( -2\beta t\right) \right) }}\right) \right\}  \notag \\
& \hspace{0.5cm}\left[ \func{Re}s>0,\beta>0,c\geqslant0,x+y\geqslant 0\right]
.   \tag{4.14}  \label{4.14}
\end{align}

\section{Some specialised results}

The above inverse Laplace transforms can be specialised by equating the
arguments (in absolute value) and/or by setting one or both of them at $0$.
This section briefly specifies four such specialisations.

First, an inverse Laplace transform for squared parabolic cylinder functions
emerges from the result in Equation (\ref{4.1}) by setting $y=x$%
\begin{align}
& L^{-1}\left\{ \Gamma\left( \tfrac{s+c}{\beta}\right) \left[ D_{-\left(
s+c\right) /\beta}\left( x\right) \right] ^{2}\right\}  \notag \\
& \hspace{0.5cm}=\dfrac{\beta\exp\left( -ct\right) }{\sqrt{1-\exp\left(
-2\beta t\right) }}\exp\left( -\dfrac{x^{2}\left( 1+\exp\left( -\beta
t\right) \right) ^{2}}{2\left( 1-\exp\left( -2\beta t\right) \right) }\right)
\notag \\
& \hspace{0.5cm}\left[ \func{Re}s>0,\beta>0,c\geqslant0,x\geqslant 0\right]
.   \tag{5.1}  \label{5.1}
\end{align}
\noindent Second, plugging $y=-x$ into the inverse transform (\ref{4.7})
gives%
\begin{align}
& L^{-1}\left\{ \Gamma\left( \tfrac{s+c}{\beta}\right) D_{-\left( s+c\right)
/\beta}\left( x\right) D_{1-\left( s+c\right) /\beta}\left( -x\right) -%
\tfrac{\sqrt{\pi}}{\sqrt{2}}\right\}  \notag \\
& \hspace{0.5cm}=\dfrac{-\beta x\left( 1-\exp\left[ -\beta t\right] \right)
\exp\left( -ct\right) }{\left( 1-\exp\left( -2\beta t\right) \right) ^{\frac{%
3}{2}}}\exp\left( -\dfrac{x^{2}\left( 1-\exp\left( -\beta t\right) \right)
^{2}}{2\left( 1-\exp\left( -2\beta t\right) \right) }\right)  \notag \\
& \hspace{0.5cm}\left[ \func{Re}s>0,\beta>0,c\geqslant0\right] .   \tag{5.2}
\label{5.2}
\end{align}
\noindent Note that the additional factor $\tfrac{\sqrt{\pi}}{\sqrt{2}}$ is
to be included irrespective of the value of the argument as the condition $%
x+y=0$ for this factor in Equation (\ref{4.7}) always holds. Also, the
inverse transform (\ref{5.2}) now applies to all real values for the
argument. As required, both sides in Equation (\ref{5.2}) vanish when
additionally restricting $x$ to $0$ as can be seen from property (\ref{pcf0}%
) and the duplication formula (\ref{gam2}).

Third, the results in the previous section generate a large body of
expressions for single parabolic cylinder functions. For instance, using $%
y=0 $ in Equation (\ref{4.1}) gives%
\begin{align}
& L^{-1}\left\{ 2^{(s+c)/\left( 2\beta\right) }\Gamma\left( \tfrac {s+c}{%
2\beta}\right) D_{-\left( s+c\right) /\beta}\left( x\right) \right\}  \notag
\\
& \hspace{0.5cm}=\dfrac{2\beta\exp\left( -ct\right) }{\sqrt{1-\exp\left(
-2\beta t\right) }}\exp\left( -\dfrac{x^{2}\left( 1+\exp\left( -2\beta
t\right) \right) }{4\left( 1-\exp\left( -2\beta t\right) \right) }\right) 
\notag \\
& \hspace{0.5cm}\left[ \func{Re}s>0,\beta>0,c\geqslant0,x\geqslant 0\right]
.   \tag{5.3}  \label{5.3}
\end{align}
\noindent The result in Equation (\ref{5.3}) complements extant inverse
transforms for single parabolic cylinder functions that are documented in Erd%
\'{e}lyi et al. \cite{emoti154} and Prudnikov, Brychkov and Marichev \cite%
{pbm592}. For instance, the Laplace transform in Equation ($5.18.9$) in Erd%
\'{e}lyi et al. \cite{emoti154} is $\overline{f}\left( s\right)
=2^{s+v}\Gamma\left( s+v\right) D_{-2s}\left( x\right) $ and its original
function is given by $f\left( t\right) =\exp\left( \tfrac{1}{2}t\right) $ $%
\left( \exp\left( t\right) -1\right) ^{-v-\frac{1}{2}}$ $\exp\left( -\tfrac{%
x^{2}\exp\left( -t\right) }{4\left( 1-\exp\left( -t\right) \right) }\right) $
$D_{2v}\left( \tfrac{x}{\sqrt{1-\exp\left( -t\right) }}\right) $. The
Laplace transform in (\ref{5.3}) thus offers a more flexible specification
for the order in the parabolic cylinder function. The Laplace transform ($%
5.18.9$), on the contrary, does not impose the tight connection between the
argument in the gamma function and the order in the parabolic cylinder
function that is present in Equation (\ref{5.3}). Note that both
specifications are identical when employing $c=0$ and $\beta=\frac{1}{2}$ in
(\ref{5.3}) and specialising ($5.18.9$) for $v=0$ and using the property (%
\ref{d0}).

Finally, setting both arguments to $0$ yields inverse Laplace transforms for
ratios of gamma functions. For instance, plugging $x=y=0$ into Equation (\ref%
{4.12}) gives%
\begin{align}
& L^{-1}\left\{ \dfrac{\left( s+c\right) \Gamma\left( \tfrac{s+c}{2\beta }%
\right) }{\left( s+c+\beta\right) \Gamma\left( \tfrac{s+c+\beta}{2\beta }%
\right) }\right\} =\dfrac{2\beta\exp\left( -\left( 2\beta+c\right) t\right) 
}{\sqrt{\pi\left( 1-\exp\left( -2\beta t\right) \right) }}  \notag \\
& \hspace{0.5cm}\left[ \func{Re}s>0,\beta>0,c\geqslant0\right] .   \tag{5.4}
\label{5.4}
\end{align}
\noindent Erd\'{e}lyi et al. \cite{emoti154} and Prudnikov, Brychkov and
Marichev \cite{pbm592} list a large number of inverse transforms for ratios
of gamma functions. The inverse transforms that can be obtained from the
results in this paper are complementary as the expressions in the literature
multiply the Laplace-transform parameter either by $1$ or $2$ (see for
instance Sections 3.1.1. and 3.1.2 in Prudnikov, Brychkov and Marichev \cite%
{pbm592}), but at the same time allow for a more flexible connection between
the arguments in the various gamma functions.

\setlength{\baselineskip}{0.75\baselineskip}

\newpage

\small%

\captionsetup[table]{labelformat=simple,justification=raggedright,singlelinecheck=false}

\begin{longtable}{l l}
\caption[]{ Inverse transforms of Laplace transforms that contain products
of two parabolic cylinder functions.* }{\label{long}}\\
\hline\hline\multicolumn{1}{l}{$\overline{f}\left( s\right) =\dint\limits
_{0}^{+\infty}\exp\left( -st\right)\ f\left( t\right) \ dt$} &
\multicolumn{1}{l}{$L^{-1}\left\{ \overline{f}\left( s\right) \right\}
=f\left( t\right)$\medskip} \\
\hline\hline\endfirsthead\caption[]{continued*}\\
\hline\hline\multicolumn{1}{l}{$\overline{f}\left( s\right) =\dint\limits
_{0}^{+\infty}\exp\left( -st\right) \ f\left( t\right) \ dt$} &
\multicolumn{1}{l}{$L^{-1}\left\{ \overline{f}\left( s\right) \right\}
=f\left( t\right)$}\medskip\\
\hline\hline\endhead\\
\hline\hline\\
{*} $\func{Re}s>0,\beta>0,c\geqslant0,x+y\geqslant0$.
\endfoot\hline\hline\\
{*} $\func{Re}s>0,\beta>0,c\geqslant0,x+y\geqslant0$.
\endlastfoot\\
1. $\Gamma\left( \dfrac{s+c}{\beta}\right) D_{1-\left( s+c\right)
/\beta}\left( x\right) D_{-1-\left( s+c\right) /\beta}\left( y\right
) $ & $\exp\left( \dfrac{y^{2}-x^{2}}{4}\right) \left\{ \dfrac{\beta\exp\left(
-\left( \beta+c\right) t\right) }{\sqrt{1-\exp\left( -2\beta t\right) }}\medskip\right. $ \\
& $\hspace{0.75cm}\times\exp\left( -\dfrac{\left( y+x\exp\left( -\beta
t\right) \right) ^{2}}{2\left( 1-\exp\left( -2\beta t\right) \right) }\right) \medskip$ \\
& $\left. +\dfrac{x\beta\sqrt{\pi}\exp\left( -ct\right) }{\sqrt{2}}\text
{erfc}\left( \dfrac{y+x\exp\left( -\beta t\right) }{\sqrt{2\left( 1-\exp
\left( -2\beta t\right) \right) }}\right) \right\} \bigskip$ \\
2. $\Gamma\left[ \dfrac{s+c}{\beta}\right] D_{-\left( s+c\right) /\beta}\left( x\right) D_{1-\left( s+c\right) /\beta}\left( y\right) $ & $\dfrac
{\beta\exp\left( -ct\right) }{\left( 1-\exp\left( -2\beta t\right) \right)
^{\frac{3}{2}}}\left( y+x\exp\left[ -\beta t\right] \right) \exp\left(
\dfrac{y^{2}-x^{2}}{4}\right) \medskip$ \\
$\hspace{1cm}-\left. \dfrac{\sqrt{\pi}}{\sqrt{2}}\right\vert_{x+y=0}$ & $\hspace{0.75cm}\times\exp\left( -\dfrac{\left( y+x\exp\left( -\beta
t\right) \right) ^{2}}{2\left( 1-\exp\left( -2\beta t\right) \right) }\right) \bigskip$ \\
3. $\Gamma\left( \dfrac{s+c}{\beta}\right) D_{-\left( s+c\right) /\beta}\left( x\right) D_{-\left( s+c\right) /\beta}\left( y\right) $ & $\dfrac
{\beta\exp\left( -ct\right) }{\sqrt{1-\exp\left( -2\beta t\right) }}\exp
\left( \dfrac{y^{2}-x^{2}}{4}\right) \medskip$ \\
& $\hspace{0.75cm}\times\exp\left( -\dfrac{\left( y+x\exp\left( -\beta
t\right) \right) ^{2}}{2\left( 1-\exp\left( -2\beta t\right) \right) }\right) \bigskip$ \\
4. $\Gamma\left( \dfrac{s+c}{\beta}\right) D_{-\left( s+c\right) /\beta}\left( x\right) D_{-1-\left( s+c\right) /\beta}\left( y\right) $ & $\dfrac
{\beta\sqrt{\pi}\exp\left( -ct\right) }{\sqrt{2}}\exp\left( \dfrac{y^{2}-x^{2}}{4}\right) \medskip$ \\
& $\hspace{0.75cm}\times$erfc$\left( \dfrac{y+x\exp\left( -\beta t\right)
}{\sqrt{2\left( 1-\exp\left( -2\beta t\right) \right) }}\right) \bigskip$
\\
5. $s\Gamma\left( \dfrac{s+c}{\beta}\right) D_{-\left( s+c\right)
/\beta}\left( x\right) D_{-1-\left( s+c\right) /\beta}\left( y\right
) $ & $\exp\left( \dfrac{y^{2}-x^{2}}{4}\right) \left\{ \dfrac{\beta^{2}\exp\left( -\left( \beta+c\right) t\right) }{\left( 1-\exp\left( -2\beta
t\right) \right) ^{\frac{3}{2}}}\right. \medskip$ \\
$\hspace{1cm}-\left. \dfrac{\beta\sqrt{\pi}}{\sqrt{2}}\right\vert_{x+y=0}$
& $\hspace{0.75cm}\times\left( y\exp\left[ -\beta t\right] +x\right) \exp
\left( -\dfrac{\left( y+x\exp\left( -\beta t\right) \right) ^{2}}{2\left(
1-\exp\left( -2\beta t\right) \right) }\right) \medskip$ \\
& $\left. -\dfrac{c\beta\sqrt{\pi}\exp\left( -ct\right) }{\sqrt{2}}\text
{erfc}\left( \dfrac{y+x\exp\left( -\beta t\right) }{\sqrt{2\left( 1-\exp
\left( -2\beta t\right) \right) }}\right) \right\} \bigskip$ \\
6. $\Gamma\left( \dfrac{s+c}{\beta}\right) D_{-\left( s+c\right) /\beta}\left( x\right) D_{-2-\left( s+c\right) /\beta}\left( y\right) $ & $\exp
\left( \dfrac{y^{2}-x^{2}}{4}\right) \left\{ \beta\sqrt{1-\exp\left(
-2\beta t\right) }\exp\left( -ct\right) \right. \medskip$ \\
& $\hspace{0.75cm}\times\exp\left( -\dfrac{\left( y+x\exp\left( -\beta
t\right) \right) ^{2}}{2\left( 1-\exp\left( -2\beta t\right) \right) }\right) \medskip$ \\
& $-\dfrac{\left( x\exp\left( -\beta t\right) +y\right) \beta\sqrt{\pi}\exp\left( -ct\right) }{\sqrt{2}}\medskip$ \\
& $\hspace{0.75cm}\times\left. \text{erfc}\left( \dfrac{y+x\exp\left(
-\beta t\right) }{\sqrt{2\left( 1-\exp\left( -2\beta t\right) \right) }}\right) \right\} \bigskip$ \\
7. $\frac{s+c}{\beta}\Gamma\left( \dfrac{s+c}{\beta}\right) D_{-\left(
s+c\right) /\beta}\left( x\right) D_{-2-\left( s+c\right) /\beta}\left(
y\right) $ & $\exp\left( \dfrac{y^{2}-x^{2}}{4}\right) \left\{ \dfrac
{\beta\exp\left( -\left( 2\beta+c\right) t\right) }{\sqrt{1-\exp\left( -2\beta
t\right) }}\right. \medskip$ \\
& $\hspace{0.75cm}\times\exp\left( -\dfrac{\left( y+x\exp\left( -\beta
t\right) \right) ^{2}}{2\left( 1-\exp\left( -2\beta t\right) \right) }\right) \medskip$ \\
& $\left. +\dfrac{x\beta\sqrt{\pi}\exp\left( -\left( \beta+c\right)
t\right) }{\sqrt{2}}\medskip\text{erfc}\left( \dfrac{y+x\exp\left( -\beta
t\right) }{\sqrt{2\left( 1-\exp\left( -2\beta t\right) \right) }}\right)
\right\} \bigskip$ \\
8. $\frac{s+\beta+c}{\beta}\Gamma\left( \dfrac{s+c}{\beta}\right)
D_{-\left( s+c\right) /\beta}\left( x\right) D_{-2-\left( s+c\right) /\beta
}\left( y\right) $ & $\exp\left( \dfrac{y^{2}-x^{2}}{4}\right) \left\{
\dfrac{\beta\exp\left( -ct\right) }{\sqrt{1-\exp\left( -2\beta t\right) }}\right. \medskip$ \\
& $\hspace{0.75cm}\times\exp\left( -\dfrac{\left( y+x\exp\left( -\beta
t\right) \right) ^{2}}{2\left( 1-\exp\left( -2\beta t\right) \right) }\right) \medskip$ \\
& $\left. -\dfrac{y\beta\sqrt{\pi}\exp\left( -ct\right) }{\sqrt{2}}\text
{erfc}\left( \dfrac{y+x\exp\left( -\beta t\right) }{\sqrt{2\left( 1-\exp
\left( -2\beta t\right) \right) }}\right) \right\} \bigskip$ \\
9. $\frac{s+2\beta+c}{\beta}\Gamma\left( \dfrac{s+c}{\beta}\right)
D_{-\left( s+c\right) /\beta}\left( x\right) D_{-3-\left( s+c\right) /\beta
}\left( y\right) $ & $\exp\left( \dfrac{y^{2}-x^{2}}{4}\right) \left\{
-y\beta\sqrt{1-\exp\left( -2\beta t\right) }\right. \medskip$ \\
& $\hspace{0.75cm}\times\exp\left( -ct\right) \exp\left( -\dfrac{\left(
y+x\exp\left( -\beta t\right) \right) ^{2}}{2\left( 1-\exp\left( -2\beta
t\right) \right) }\right) \medskip$ \\
& $+\dfrac{\left( y\left( x\exp\left( -\beta t\right) +y\right) +1\right)
\beta\sqrt{\pi}\exp\left( -ct\right) }{\sqrt{2}}\medskip$ \\
& $\hspace{0.75cm}\times\left. \text{erfc}\left( \dfrac{y+x\exp\left(
-\beta t\right) }{\sqrt{2\left( 1-\exp\left( -2\beta t\right) \right) }}\right) \right\} \bigskip$ \\
10. $\frac{s+c}{\beta}\Gamma\left( \dfrac{s+c}{\beta}\right)
D_{-1-\left( s+c\right) /\beta}\left( x\right) D_{-1-\left( s+c\right)
/\beta}\left( y\right) $ & $\dfrac{\beta\exp\left( -\left( \beta
+c\right) t\right) }{\sqrt{1-\exp\left( -2\beta t\right) }}\exp\left(
\dfrac{y^{2}-x^{2}}{4}\right) \medskip$ \\
& $\hspace{0.75cm}\times\exp\left( -\dfrac{\left( y+x\exp\left( -\beta
t\right) \right) ^{2}}{2\left( 1-\exp\left( -2\beta t\right) \right) }\right) \bigskip$ \\
\end{longtable}%


\begin{thebibliography}{99}
\bibitem{kcw96} Kalmykov YP, Coffey WT, Waldron JT. Exact analytic solution
for the correlation time of a Brownian particle in a doublewell potential
from the Langevin equation. J. Chem. Phys. 1996;105:2112--2118.

\bibitem{dfg92} deLyra JL, Foong SK, Gallivant TE. Finite lattice systems
with true critical behavior. Phys. Rev. D. 1992;46:1643--1657.

\bibitem{zm07} Zaqarashvili, TV, Murawski, K. Torsional oscillations of
longitudinally inhomogeneous coronal loops. Astron. Astrophys.
2007;470:353--357.

\bibitem{c43} Chandrasekhar S. Dynamical friction. I. General
considerations: the coefficient of dynamical friction. Astrophys. J.
1943;97:255--262.

\bibitem{g67} Gluss B. A model for neuron firing with exponential decay of
potential resulting in diffusion equations for probability density. Bull.
Math. Biophys. 1967;29:233--243.

\bibitem{v77} Vasicek O. An equilibrium characterization of the term
structure. J. Fin. Ec. 1977;5:177--188.

\bibitem{d42} Doob JL. The Brownian movement and stochastic equations. Ann.
Math. 1942;43:351--369.

\bibitem{cm72} Cox DR, Miller HD. The Theory of Stochastic Processes.
London: Chapman \& Hall; 1972.

\bibitem{r89} Risken H. The Fokker--Planck Equation. Methods of Solution and
Applications. 2nd ed. Berlin: Springer; 1989.

\bibitem{emoti154} Erd\'{e}lyi A, Magnus W, Oberhettinger F, Tricomi FG.
Table of Integral Transforms. Vol. 1. New York (NY): McGraw-Hill; 1954.

\bibitem{pbm592} Prudnikov AP, Brychkov YA, Marichev OI. Integrals and
series. Volume 5. Inverse Laplace transforms. New York (NY): Gordon and
Breach; 1992.

\bibitem{db07} Debnath L, Bhatta D. Integral transforms and their
applications. Boca Raton (FL): Chapman \& Hall/CRC Press; 2007.

\bibitem{emoth253} Erd\'{e}lyi A, Magnus W, Oberhettinger F, Tricomi FG.
Higher Transcendental Functions. Vol. 2. New York (NY): McGraw-Hill; 1953.

\bibitem{as72} Abramowitz M, Stegun IA. Handbook of Mathematical Functions
with Formulas, Graphs, and Mathematical Tables. New York (NY): Dover
Publications; 1972.
\end{thebibliography}
\end{document}